\documentclass[prb,twocolumn,showpacs,preprintnumbers,amsmath,amssymb]{revtex4}
\usepackage{graphicx}
\usepackage{longtable}
\usepackage{dcolumn}
\usepackage{bm}

\def\sgn{\mathop{\rm sgn}\nolimits}

\begin{document}

\title{Wannier-function approach to spin excitations in solids}

\author{Ersoy
\c{S}a\c{s}{\i}o\u{g}lu$^{1}$}\email{e.sasioglu@fz-juelich.de}
\author{Arno Schindlmayr$^{2}$}
\author{Christoph Friedrich$^{1}$}
\author{Frank Freimuth$^{1}$}
\author{Stefan  Bl\"{u}gel$^{1}$}

\affiliation{$^{1}$Institut f\"{u}r Festk\"{o}rperforschung and
Institute for Advanced Simulation,
Forschungszentrum J\"{u}lich, 52425 J\"{u}lich, Germany \\
$^{2}$Department Physik, Universit\"{a}t Paderborn, 33095
Paderborn, Germany}

\date{\today}

\begin{abstract}

We present a computational scheme to study spin excitations in
magnetic materials from first principles. The central quantity is
the transverse spin susceptibility, from which the complete
excitation spectrum, including single-particle spin-flip Stoner
excitations and collective spin-wave modes, can be obtained. The
susceptibility is derived from many-body perturbation theory and
includes dynamic correlation through a summation over ladder
diagrams that describe the coupling of electrons and holes with
opposite spins. In contrast to earlier studies, we do not use a
model potential with adjustable parameters for the electron-hole
interaction but employ the random-phase approximation. To reduce
the numerical cost for the calculation of the four-point
scattering matrix we perform a projection onto maximally localized
Wannier functions, which allows us to truncate the matrix
efficiently by exploiting the short spatial range of electronic
correlation in the partially filled \textit{d} or \textit{f}
orbitals. Our implementation is based on the full-potential
linearized augmented-plane-wave (FLAPW) method. Starting from a
ground-state calculation within the local-spin-density
approximation (LSDA), we first analyze the matrix elements of the
screened Coulomb potential in the Wannier basis for the
3\textit{d} transition-metal series. In particular, we discuss the
differences between a constrained nonmagnetic and a proper
spin-polarized treatment for the ferromagnets Fe, Co, and Ni. The
spectrum of single-particle and collective spin excitations in fcc
Ni is then studied in detail. The calculated spin-wave dispersion
is in good overall agreement with experimental data and contains
both an acoustic and an optical branch for intermediate wave
vectors along the $[1\,0\,0]$ direction. In addition, we find
evidence for a similar double-peak structure in the spectral
function along the $[1\,1\,1]$ direction. To investigate the
influence of static correlation we finally consider LSDA+$U$ as an
alternative starting point and show that, together with an
improved description of the Fermi surface, it yields a more
accurate quantitative value for the spin-wave stiffness constant,
which is overestimated in the LSDA.

\end{abstract}

\pacs{71.15.Qe, 71.45.Gm, 75.30.Ds, 71.20.Be}

\maketitle

\section{introduction}
\label{introduction}

Spin excitations in solids are of fundamental interest for a wide
variety of phenomena. For example, in magnetic materials at low
temperatures collective spin excitations, so-called spin waves or
magnons,\cite{Moriya} leave a mark on the transport, dynamical,
and thermodynamical properties. The spin waves contribute to the
specific heat with a $T^{3/2}$ term in addition to the $T^3$ term
from phonon excitations. The low-temperature behavior of the
magnetization in three-dimensional magnetic solids is also
dominated by spin-wave excitations: In ferromagnets the
magnetization drops as $T^{3/2}$, while the sublattice
magnetization in antiferromagnets  obeys a $T^2$ law.\cite{Kobler}
In low-dimensional systems spin-wave excitations even destroy the
long-range magnetic order completely at any finite temperature $T$
in the absence of a magnetic anisotropy.\cite{Mermin} As the
temperature increases, additional single-particle spin-flip
processes, the so-called Stoner excitations, take place, which
further contribute to the temperature variation of the
magnetization and cause a damping of the spin-wave modes. Spin
waves can also couple to charge excitations around the Fermi
level. Such interactions lead to a pronounced energy
renormalization in the quasiparticle band
dispersion,\cite{Hofmann:09} or control the spin-dependent
inelastic mean free path of hot electrons in
ferromagnets.\cite{Hong_1,Hong_2,Zhukov_1,Zhukov_2} Another
interesting phenomenon is high-temperature superconductivity, in
which spin waves have been proposed as a possible mediator for the
attractive electron-electron interaction.\cite{Dagotto,Scalapino}
This scenario was recently confirmed by infrared spectroscopy and
neutron scattering measurements for the high-temperature
superconductor YBa$_2$Cu$_3$O$_{6.92}$.\cite{Carbotte,Dahm}

Spin excitations in solids are also of central interest in the
field of spintronics. The writing process of magnetic information
in a giant magneto resistance or tunnel magneto resistance device
is closely related to the rotation of the magnetization, a process
generating and radiating spin waves at all wave lengths, whose
damping rate is an important parameter determining the writing
time.\cite{Choi} Spin waves are also generated during the reading
process, as hot electrons impinge at the interfaces between the
insulating barrier and the magnetic electrodes of magnetic tunnel
junctions, which causes a reduction of the magneto
resistance.\cite{Farle}

The properties and physics of spin waves evidently comprise an
unusually rich area of research. A lot of information about the
spin dynamics in solids can be obtained from the magnetic response
function (or dynamical spin susceptibility). The spectrum of
magnetic excitations corresponds to the poles of the response
function and can be directly compared with experiments like
inelastic neutron scattering. In this way it provides insight into
the nature of the exchange coupling and the complex magnetic
order. The magnetic response function is thus a central quantity
for the theoretical description of magnetic materials.

So far most theoretical studies of magnetic excitations in solids
were based on an adiabatic treatment of the spin degrees of
freedom in which the slow motion of the magnetic moments and the
fast motion of the electrons are separated by mapping the complex
itinerant electron problem onto the classical Heisenberg
Hamiltonian with exchange parameters obtained from constrained
density-functional theory (DFT)
calculations.\cite{Kubler,Sandratskii,Nordstrom,Johansson,Halilov,Schilfgaarde,Fahnle,Pajda}
Within this approach the spin-wave excitations can be calculated
efficiently, whereas the single-particle Stoner excitations are
neglected. Furthermore, the spin-wave life times are not
accessible. From a fundamental point of view, the Heisenberg model
is justified only for local-moment systems like insulators and
rare earths, which possess well-defined spin-wave modes over the
entire Brillouin zone, so that the adiabatic approximation is
reliable. For itinerant-electron magnets the adiabatic
approximation yields reasonable results only in the
long-wave-length region, i.e., for small wave vectors. For short
wave lengths the discrepancy with experiments can be large,
however. For example, the multiple branches in the spin-wave
dispersion of 3\emph{d} ferromagnets cannot be
captured.\cite{Cooke_1,Cooke_2}

First-principles calculations of the magnetic response function
using realistic energy bands and wave functions are very rare.
Initial attempts by Cooke \emph{et
al.}\cite{Cooke_1,Cooke_2,Cooke_3,Cooke_4} within the framework of
many-body perturbation theory (MBPT) were based on a tight-binding
description of the electronic energy bands. These authors studied
the spin-wave dispersion of 3\emph{d} ferromagnets and obtained
reasonable agreement with experiments; in particular, the optical
branch in the spin-wave dispersion of fcc Ni was correctly
described.\cite{Cooke_3} Using a similar approach, Mills and
co-workers\cite{Mills_1,Mills_2,Mills_3,Mills_4} carried out
extensive calculations to explore the spin dynamics in ultrathin
ferromagnetic films on nonmagnetic substrates. Recently more
realistic treatments of the spin-wave spectra in 3\emph{d}
ferromagnets were reported by Savrasov\cite{Savrasov_SW} and
Buczek \emph{et al.}\cite{Pavel} within time-dependent
density-functional theory (TDDFT) and by Karlsson and Aryasetiawan
within MBPT.\cite{Karlsson_1} However, in the latter work the
authors used a simplified  model potential with an adjustable
parameter to estimate the matrix elements of the screened Coulomb
interaction. Under these circumstances both TDDFT and MBPT appear
to give similar results for the spin-wave dispersions of Fe and
Ni. The results for Fe are in good agreement with available
experimental data, while for Ni the optical branch is too high in
energy, which was attributed to the overestimation of the exchange
splitting of Ni within the underlying LSDA.\cite{Karlsson_1}

The aim of the present work is to develop a practical
computational scheme to study excitation spectra of magnetic
materials from first principles. The magnetic response function is
calculated within a many-body context following the formalism
given in Ref.\,\onlinecite{Aryasetiawan}. To study collective
spin-wave excitations we include vertex corrections in the form of
ladder diagrams, which describe the coupling of electrons and
holes with opposite spins via the screened Coulomb interaction. In
analogy to the $T$-matrix that describes the particle-particle
scattering channel, here we use the same term for the
electron-hole channel in agreement with the definition of
Strinati.\cite{Strinati} In contrast to earlier treatments, the
matrix elements of the screened Coulomb potential are calculated
entirely from first principles. In order to reduce the numerical
cost for the calculation of the four-point $T$-matrix we exploit a
transformation to maximally localized Wannier functions (MLWFs),
which provide a more efficient basis to study local correlations
than extended Bloch
states.\cite{Wannier_1,Wannier_2,Wannier_3,Wannier_4,Wannier_5,Wannier_6,Wannier_7,LC_1,LC_2,LC_3,LC_4}
This use of localized orbitals makes our scheme very efficient for
complex magnetic materials with many atoms per unit cell. Our
implementation is based on the FLAPW method. In the following we
first calculate the matrix elements of the Coulomb potential for
the 3\emph{d} transition-metal series in the Wannier basis and
perform extensive convergence tests. The magnetic excitations in
fcc Ni are then studied in detail based on the LSDA and LSDA+$U$
methods. We find that both approaches yield qualitatively similar
results for the spin-wave spectra and overall dispersion. However,
the static correlation effects seem to be important for the
spin-wave stiffness constant, which is overestimated within LSDA.
In contrast to some previous theoretical
studies,\cite{Cooke_3,Karlsson_1} our calculations clearly
indicate the existence of an optical branch in the spin-wave
dispersion curve of Ni along the $[1\,1\,1]$ in addition to that
along the $[1\,0\,0]$  direction in the Brillouin zone. In
general, the obtained results are in good agreement with available
experimental data.

This paper is organized as follows. In Sec.\,\ref{sectionII} we
describe the computational method. Section\,\ref{sectionIII}
contains the results for the matrix elements of the screened
Coulomb potential for the 3\emph{d} transition metals. In
Sec.\,\ref{sectionIV} we present results for the magnetic
excitations in fcc Ni together with a detailed discussion. In
Sec.\,\ref{sectionV} we summarize our conclusions and give an
outlook. Unless otherwise indicated, Hartree atomic units are used
throughout.

\section{Computational Method}
\label{sectionII}

\subsection{Magnetic response function}

The time-ordered magnetic response function (or dynamical spin
susceptibility)  is given in real space by the correlation
function
\begin{equation}
R^{ij}(1,2)= -i\langle \mathcal{T}
[\hat{\sigma}^{i}(1),\hat{\sigma}^{j}(2)]\rangle\:,
\label{response_function0}
\end{equation}
where $\mathcal{T}$ is the time-ordering operator and
$\hat{\sigma}^{i}(1)$ are the spin-density operators with $i \in
\{x,y,z,-,+\}$, where $-$ and $+$ correspond to the spin annihilation
($\hat{\sigma}^{-}=\hat{\sigma}^{x}-i\hat{\sigma}^{y}$) and
creation ($\hat{\sigma}^{+}=\hat{\sigma}^{x}+i\hat{\sigma}^{y}$)
operator, respectively. For simplicity we use the short-hand
notation $1=(\mathbf{r}_1,t_1)$. The expectation value of
$\hat{\sigma}^{i}(1)$  with respect to the many-body ground
state is given by
\begin{equation}
\langle \hat{\sigma}^{i}(1)\rangle=
-i\sum_{\alpha,\beta}\sigma^{i}_{\beta\alpha}
G_{\alpha\beta}(1,1^{+})
\label{spin_density}
\end{equation}
with the  Pauli spin matrices $\sigma^{i}$, the single-particle
Green function $G$, and the spin indices $\alpha$ and $\beta$. The
notation $1^{+}$ indicates that the time variable is increased by
an infinitesimal to ensure the proper time ordering $t_{1}^{+} > t_{1}$.
The magnetic response function can be obtained from the spin
density by the functional derivative
\begin{equation}
R^{ij}(1,2)=
\frac{\delta\langle \hat{\sigma}^{i}(1)\rangle} {\delta
B_{j}(2)}\:,
\label{response_function1}
\end{equation}
where $i$ and $j$ correspond to the components of the
magnetization and the magnetic-field vector, respectively. The
latter incorporates a factor $g\mu_{\mathrm{B}}/2$, where the
$\mu_{\mathrm{B}}$ denotes the Bohr magneton and $g$ the electron
$g$-factor, so that the Zeeman term in the Hamiltonian operator
takes the form $+ \mathbf{B}\cdot \mathbf{\sigma}$.

\subsection{$T$-matrix approximation}

\begin{figure}[t]
\begin{center}
\includegraphics[scale=0.5]{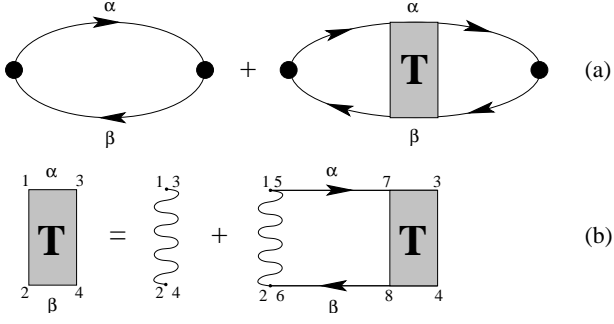}
\end{center}
\vspace*{-0.4cm} \caption{ Diagrammatic representation of (a) the
magnetic response function and (b) the $T$-matrix.}
\label{response}
\end{figure}

The single-particle Green function in Eq.\,(\ref{spin_density})
is given  by the Dyson equation
\begin{eqnarray}
G_{\alpha\beta}(1,2)&=&G^{0}_{\alpha}(1,2)\delta_{\alpha\beta} +
\sum_{\gamma}\iint G^{0}_{\alpha}(1,3) \nonumber\\ && \times
\Sigma_{\alpha\gamma}(3,4) G_{\gamma\beta}(4,2)\:d3\:d4\:,
\label{Dyson_equation}
\end{eqnarray}
where $G^{0}$ is the spin-diagonal Green function of the
noninteracting Hartree system and $\Sigma$ the nonlocal dynamic
self-energy, which incorporates all exchange-correlation effects.
With the identity
\begin{eqnarray}
\lefteqn{\frac{\delta G_{\alpha\beta}(1,3)}{\delta B_{j}(2)}}
\nonumber \\ &&= -\sum_{\gamma,\delta}\iint
G_{\alpha\gamma}(1,4)\frac{\delta
G^{-1}_{\gamma\delta}(4,5)}{\delta B_{j}(2)}
G_{\delta\beta}(5,3)\:d4\:d5 \label{Identity}
\end{eqnarray}
one can rewrite the magnetic response function
(\ref{response_function1}) as
\begin{eqnarray}
\lefteqn{R^{ij}(1,2)} \nonumber \\
&=&-i\sum_{\alpha,\beta,\gamma,\delta}\sigma^{i}_{\beta\alpha} \iint
G_{\alpha\gamma}(1,3)
\bigg [\sigma^{j}_{\gamma\delta}\delta(2-3) \nonumber \\
&& \times \delta(3-4) +
\frac{\delta\Sigma_{\gamma\delta}(3,4)}{\delta B_{j}(2)}\bigg
]G_{\delta\beta}(4,1^{+})\:d3\:d4\:.
\label{response_function2}
\end{eqnarray}
In this work we use the  $GW$
approximation for the self-energy\cite{Hedin}
\begin{equation}
\Sigma_{\gamma\delta}(3,4)=iG_{\gamma\delta}(3,4)W(3,4)\:.
\label{Self-energy}
\end{equation}
The functional derivative of the self-energy with respect to the
external magnetic field is then given by
\begin{eqnarray}
\frac{\delta\Sigma_{\gamma\delta}(3,4)}{\delta B_{j}(2)}&=&
i\frac{\delta G_{\gamma\delta}(3,4)}{\delta B_{j}(2)}W(3,4)+iG_{\gamma\delta}(3,4) \nonumber \\
&& \times \frac{\delta W(3,4)}{\delta B_{j}(2)}\:.
\label{Self-energy2}
\end{eqnarray}
For systems with a collinear magnetic ground state only the first
term on the right-hand  side yields a nonzero contribution to the
magnetic response function.\cite{Aryasetiawan} Furthermore, in
this case the Green function is diagonal in spin space and can be
written as
$G_{\alpha\beta}(1,2)=G_{\alpha}(1,2)\delta_{\alpha\beta}$.
The dynamically screened Coulomb potential $W(3,4)$ is given  by
\begin{equation}
W(3,4)= v(3,4) + \iint v(3,5)P(5,6)W(6,4)\:d5\:d6\:,
\label{Screened_W}
\end{equation}
where $v(3,4)=\delta(t_3-t_4)/|\mathbf{r}_3-\mathbf{r}_4|$ is
the bare Coulomb potential and $P(5,6)$ the polarizability in
the random-phase approximation (RPA). The latter is expressed by
\begin{equation}
P(5,6)=-\sum_{\alpha}K^{\alpha\alpha}(5,6;6,5)\:,
\label{Polarization}
\end{equation}
where the kernel $K$ is defined as
\begin{equation}
K^{\alpha\beta}(1,3;4,2)=iG_{\alpha}(1,3)G_{\beta}(4,2^{+})\:.
\label{Kernel_K}
\end{equation}
After collecting all terms we obtain

\begin{eqnarray}
R^{ij}(1,2)&=& -\sum_{\alpha,\beta}\sigma^{i}_{\beta\alpha}
\sigma^{j}_{\alpha\beta}\big[K^{\alpha\beta}(1,2;2,1) \nonumber  \\
&&+ L^{\alpha\beta}(1,2;2,1) \big]
\label{response_function3}
\end{eqnarray}
for the magnetic response function. The second contribution is
given by
\begin{eqnarray}
L^{\alpha\beta}(1,2;2,1)&=& \iiiint
K^{\alpha\beta}(1,3;4,1)T^{\alpha\beta}(3,5;6,4) \nonumber \\ &&
\times K^{\alpha\beta}(5,2;2,6)\:d3\:d4\:d5\:d6\:,
\label{correlation_part}
\end{eqnarray}
where the $T$-matrix obeys the Bethe-Salpeter equation
\begin{eqnarray}
\lefteqn{T^{\alpha\beta}(1,3;4,2)} \nonumber \\
&=& W(1,2)\delta(1-3)\delta(2-4) + W(1,2)\nonumber \\
&&\times \iint
K^{\alpha\beta}(1,5;6,2)T^{\alpha\beta}(5,3;4,6)\:d5\:d6\:.
\label{Bethe-Salpeter}
\end{eqnarray}
The first term in Eq.\,(\ref{response_function3}) represents the
response of the noninteracting system, i.e., the Kohn-Sham spin
susceptibility. The second term contains the $T$-matrix, which
describes dynamic correlation in the form of repeated scattering
events of particle-hole pairs with opposite spins and is
responsible for the occurrence of collective spin-wave
excitations. The Feynman diagrams for the magnetic response
function $R$ and the $T$-matrix are displayed in
Fig.\,\ref{response}.

For a practical evaluation of the magnetic response function we
replace the full renormalized Green function $G_{\alpha}(1,2)$ by
an appropriate mean-field approximation (LSDA, LSDA+$U$, etc.).
Moreover, we employ an instantaneous interaction of the form
$W(\mathbf{r}_1, \mathbf{r}_2; t_1-t_2 )\approx W(\mathbf{r}_1,
\mathbf{r}_2)\,\delta(t_1-t_2)$ with $W(\mathbf{r}_1,
\mathbf{r}_2) = \int_{- \infty}^{\infty} W(\mathbf{r}_1,
\mathbf{r}_2;\tau)\,d\tau$ in Eq.\,(\ref{Bethe-Salpeter}). This is
justified by the observation that the matrix elements of
$W(\mathbf{r}_1, \mathbf{r}_2)$ do not vary strongly in the
low-frequency region ($\omega < 1$ eV), in which the spin-wave
excitations occur.\cite{Jepsen,Miyake} Under these circumstances
it is sufficient to calculate the kernel (\ref{Kernel_K}) for
$t_2=t_1$ and  $t_4=t_3$. In fact, due to time translation
symmetry the resulting expression depends only on the difference
$t_1-t_3$ and is given by the Fourier transform of
\begin{eqnarray}
\lefteqn{K^{\alpha\beta}(\textbf{r}_1,\textbf{r}_3;\textbf{r}_4,\textbf{r}_2
;\omega) } \nonumber \\ &=&\sum_{\mathbf{k},m}^{\textmd{occ}}
\sum_{\mathbf{k}^{\prime},m^{\prime}}^{\textmd{unocc}} \bigg(\frac{
\varphi_{\mathbf{k}m}^{\alpha}(\textbf{r}_1)
\varphi_{\mathbf{k}m}^{\alpha*}(\textbf{r}_3)
\varphi_{\mathbf{k}^{\prime}m^{\prime}}^{\beta}(\textbf{r}_4)
\varphi_{\mathbf{k}^{\prime}m^{\prime}}^{\beta*}(\textbf{r}_2)}
{\omega+(\epsilon_{\mathbf{k}^{\prime}m^{\prime}}^{\beta}-\epsilon_{\mathbf{k}m}^{\alpha})-i\delta}
\nonumber\\ &&-\frac{
\varphi_{\mathbf{k}^{\prime}m^{\prime}}^{\alpha}(\textbf{r}_1)
\varphi_{\mathbf{k}^{\prime}m^{\prime}}^{\alpha*}(\textbf{r}_3)
\varphi_{\mathbf{k}m}^{\beta}(\textbf{r}_4)
\varphi_{\mathbf{k}m}^{\beta*}(\textbf{r}_2)}
{\omega-(\epsilon_{\mathbf{k}^{\prime}m^{\prime}}^{\alpha}-\epsilon_{\mathbf{k}m}^{\beta})+i\delta}\bigg)\:,
\label{kernel1}
\end{eqnarray}
where $\varphi_{\mathbf{k}m}^{\alpha}(\textbf{r})$ and
$\epsilon_{\mathbf{k}m}^{\alpha}$ are the LSDA (LSDA+$U$) eigenstates and
eigenvalues, respectively.

\subsection{Implementation in the Wannier basis}

In order to reduce the numerical cost for the calculation of the
four-point kernel $K$ we exploit a transformation to maximally
localized Wannier functions, which allows us to efficiently
truncate the matrix in real space. The generalized Wannier
functions $w_{n\textbf{R}}^{\alpha}(\textbf{r})$ with orbital
index $n$ and spin $\alpha$ at the site $\textbf{R}$ are defined
as Fourier transforms of the Bloch states
$\varphi_{\mathbf{k}m}^{\alpha}(\textbf{r})$ according to
\begin{eqnarray}
w_{n\textbf{R}}^{\alpha}(\textbf{r})&=&
\frac{1}{N}\sum_{\mathbf{k}}e^{-i\mathbf{k}\cdot\textbf{R}}\sum_{m}U_{mn}^{\alpha(\mathbf{k})}
\varphi_{\mathbf{k}m}^{\alpha}(\textbf{r}) \nonumber\\
&=&\frac{1}{N}\sum_{\mathbf{k}}e^{-i\mathbf{k}\cdot\textbf{R}}w_{\mathbf{k}n}^{\alpha}(\textbf{r})\:,
\label{Wannier}
\end{eqnarray}
where $N$ is the number of discrete $\mathbf{k}$ points in the
full Brillouin zone and $U_{mn}^{\alpha(\mathbf{k})}$ denote the
transformation matrices. The latter are determined by minimizing
the spread
\begin{equation}
\Omega=\sum_{n,\alpha}\big(\langle
w_{n\mathbf{0}}^{\alpha}|r^{2}|w_{n\mathbf{0}}^{\alpha} \rangle -\langle
w_{n\mathbf{0}}^{\alpha}|\textbf{r}|w_{n\mathbf{0}}^{\alpha} \rangle^2\big)\:,
\label{spread}
\end{equation}
where the sum runs over all Wannier functions. We employ the
algorithm for minimizing the spread initially proposed by Marzari
and Vanderbilt\cite{Wannier_1} for isolated groups of bands and
later extended to entangled energy bands.\cite{Wannier_2}

The matrix elements of the screened Coulomb potential in the MLWF
basis are given by
\begin{eqnarray}
\lefteqn{W^{\alpha\beta}_{n_1\textbf{R}_1n_3\textbf{R}_3;n_4\textbf{R}_4n_2\textbf{R}_2}} \nonumber \\
&=&\iint
w_{n_1\textbf{R}_1}^{\alpha*}(\textbf{r})w_{n_3\textbf{R}_3}^{\alpha}(\textbf{r})
W(\textbf{r},\textbf{r}^{\prime}) \nonumber \\
&& \times w_{n_4\textbf{R}_4}^{\beta*}(\textbf{r}^{\prime})
w_{n_2\textbf{R}_2}^{\beta}(\textbf{r}^{\prime})\: d^3r\:d^3r^{\prime}\:.
\label{Coulomb_1}
\end{eqnarray}
The screened potential $W(\textbf{r},\textbf{r}^{\prime})$ itself
is calculated within the RPA using the mixed product
basis.\cite{Mixed_Basis,Mixed_Basis_2,Friedrich} As we use only
the on-site matrix elements of  $W$, we set
$\textbf{R}_1=\textbf{R}_2=\textbf{R}_2=\textbf{R}_4$ and
eventually obtain
\begin{eqnarray}
\lefteqn{W^{\alpha\beta}_{n_1n_3;n_4n_2}}
\nonumber \\ &=&
\frac{1}{N^3}\sum_{\mathbf{k},\mathbf{q}_1,\mathbf{q}_2}
\sum_{I,J} \sum_{m_1,m_2,m_3,m_4}
U_{m_1n_1}^{\alpha(\mathbf{k}+\mathbf{q}_1)*}
U_{m_2n_2}^{\beta(\mathbf{k}+\mathbf{q}_2)}\:\:\:\:\:\nonumber \\ &&
\times U_{m_3n_3}^{\alpha(\mathbf{q}_1)}
U_{m_4n_4}^{\beta(\mathbf{q}_2)*}
 \langle \varphi_{\mathbf{k}+\mathbf{q}_1 m_1}^{\alpha}|\varphi_{\mathbf{q}_1 m_3}^{\alpha}\tilde{M}_{I\mathbf{k}}\rangle  \nonumber \\
 && \times
 \langle M_{I\mathbf{k}}|W(\textbf{r},\textbf{r}^{\prime})|M_{J\mathbf{k}}\rangle \langle \tilde{M}_{J\mathbf{k}}\varphi_{\mathbf{q}_2 m_4}^{\beta}| \varphi_{\mathbf{k}+\mathbf{q}_2 m_2}^{\beta}\rangle\:,
\label{Coulomb_2}
\end{eqnarray}
where $M_{I\mathbf{k}}$ ($\tilde{M}_{I\mathbf{k}}$) are the
biorthogonal basis functions of  the mixed product basis, which
satisfy the relation
\begin{equation}
\sum_{I,\mathbf{k}}|M_{I\mathbf{k}}\rangle \langle
\tilde{M}_{I\mathbf{k}}|=1
\end{equation}
in the Hilbert space of the wave-function products.

The next step is the calculation of the kernel $K$. The projection
of the Bloch states onto the Wannier orbitals yields
\begin{equation}
\int w_{n\textbf{R}}^{\alpha*}(\textbf{r})
\varphi_{\mathbf{k}m}^{\alpha}(\textbf{r})\:d^3r=
U_{mn}^{\alpha(\mathbf{k})*}e^{i\mathbf{k}\cdot\textbf{R}}\:.
\label{projection}
\end{equation}
If we perform a lattice Fourier transformation, then
Eq.\,(\ref{kernel1}) takes the form
\begin{eqnarray}
\lefteqn{K^{\alpha\beta}_{n_1n_3;n_4n_2} (\mathbf{q},\omega)}
\nonumber \\ &=& \frac{1}{N} \sum_{\mathbf{k}}
\sum_{m}^{\textmd{occ}} \sum_{m^{\prime}}^{\textmd{unocc}}
\bigg(\frac{U_{mn_1}^{\alpha(\mathbf{k})}
U_{mn_3}^{\alpha(\mathbf{k})*}
U_{m^{\prime}n_4}^{\beta(\mathbf{k}+\mathbf{q})}
U_{m^{\prime}n_2}^{\beta(\mathbf{k}+\mathbf{q})*}}
{\omega+(\epsilon_{\mathbf{k}+\mathbf{q}m^{\prime}}^{\beta}-\epsilon_{\mathbf{k}m}^{\alpha})-i\delta}\nonumber\\
&&-\frac{
U_{m^{\prime}n_1}^{\alpha(\mathbf{k}+\mathbf{q})*}
U_{m^{\prime}n_3}^{\alpha(\mathbf{k}+\mathbf{q})}
U_{mn_4}^{\beta(\mathbf{k})*}
U_{mn_2}^{\beta(\mathbf{k})}}
{\omega-(\epsilon_{\mathbf{k}+\mathbf{q}m^{\prime}}^{\alpha}-\epsilon_{\mathbf{k}m}^{\beta})+i\delta}\bigg)
\label{kernel_Wannier}
\end{eqnarray}
in the Wannier basis. Instead of a direct evaluation of this
expression, we first calculate the corresponding spectral function
\begin{eqnarray}
\lefteqn{S^{\alpha\beta}_{n_1n_3;n_4n_2}
(\mathbf{q},\omega)} \nonumber \\
&=&\frac{1}{N}
\sum_{\mathbf{k}}
\sum_{m}^{\textmd{occ}}
\sum_{m^{\prime}}^{\textmd{unocc}}
\bigg(U_{m^{\prime}n_1}^{\alpha(\mathbf{k}+\mathbf{q})*}
U_{m^{\prime}n_3}^{\alpha(\mathbf{k}+\mathbf{q})}
U_{mn_4}^{\beta(\mathbf{k})*}
 \nonumber\\
&&\times  U_{mn_2}^{\beta(\mathbf{k})}\delta(\omega-\epsilon_{\mathbf{k}+\mathbf{q}m^{\prime}}^{\alpha}+\epsilon_{\mathbf{k}m}^{\beta})
 -U_{mn_1}^{\alpha(\mathbf{k})}U_{mn_3}^{\alpha(\mathbf{k})*}
 \nonumber\\
&& \times  U_{m^{\prime}n_4}^{\beta(\mathbf{k}+\mathbf{q})}
U_{m^{\prime}n_2}^{\beta(\mathbf{k}+\mathbf{q})*}
\delta(\omega+\epsilon_{\mathbf{k}+\mathbf{q}m^{\prime}}^{\beta}-\epsilon_{\mathbf{k}m}^{\alpha})\bigg)\:,
\label{spectral_sunction}
\end{eqnarray}
which equals the probability distribution for spin-flip
transitions between occupied and unoccupied states with the energy
and momentum difference $\omega$ and $\mathbf{q}$. Once the
spectral function is known, we use a Hilbert transformation to
calculate the kernel
\begin{eqnarray}
K^{\alpha\beta}_{n_1n_3;n_4n_2}
(\mathbf{q},\omega)&=&-\mathcal{P}\int_{-\infty}^{\infty}\frac{S^{\alpha\beta}_{n_1n_3;n_4n_2}
(\mathbf{q},\omega^{\prime})}{\omega-\omega^{\prime}}\:d\omega^{\prime} \nonumber \\
&& +\: i\pi S^{\alpha\beta}_{n_1n_3;n_4n_2}
(\mathbf{q},\omega)\sgn(\omega)\:,
\label{Hilbert_transform}
\end{eqnarray}
where $\mathcal{P}$ indicates the Cauchy principal value.

With the kernel $K$ and the screened Coulomb potential $W$ we can
construct the $T$-matrix according to Eq.\,(\ref{Bethe-Salpeter}),
which in the MLWF basis takes the form
\begin{eqnarray}
\lefteqn{T^{\alpha\beta}_{n_1n_3;n_4n_2}
(\mathbf{q},\omega)} \nonumber \\
&=& W^{\alpha\beta}_{n_1n_3;n_4n_2}+
\sum_{n_5,n_6,n_7,n_8}W^{\alpha\beta}_{n_1n_5;n_6n_2}K^{\alpha\beta}_{n_5n_7;n_8n_6}(\mathbf{q},\omega)
\nonumber \\ && \times
 T^{\alpha\beta}_{n_7n_3;n_4n_8}
(\mathbf{q},\omega) \label{$T$-matrix2}
\end{eqnarray}
and can be solved by a matrix inversion for a set of $\mathbf{q}$
and $\omega$ values. Finally, the magnetic response function is
given by
\begin{eqnarray}
R^{ij}(\mathbf{q},\omega)&=&-\sum_{\alpha,\beta}\sum_{\mathbf{k}}\sum_{n_1,n_2,n_3,n_4}\sigma^{i}_{\beta\alpha}
\sigma^{j}_{\alpha\beta}
\nonumber \\
&&\times \langle
\mathbf{q}|\tilde{w}^{\alpha}_{n_1\mathbf{k}+\mathbf{q}}\tilde{w}^{\beta*}_{n_2\mathbf{k}}\rangle
\big[K_{n_1n_3;n_4n_2}^{\alpha\beta}(\mathbf{q},\omega) \nonumber
\\ && + L_{n_1n_3;n_4n_2}^{\alpha\beta}(\mathbf{q},\omega)\big]
\langle \tilde{w}^{\alpha}_{n_3\mathbf{k}+
\mathbf{q}}\tilde{w}^{\beta*}_{n_4\mathbf{k}}| \mathbf{q} \rangle
\:\:\:\:\:\:\:\:\:\: \label{response_function4}
\end{eqnarray}
with
\begin{equation}
\langle\mathbf{q}|\tilde{w}^{\alpha}_{n_1\mathbf{k}+\mathbf{q}}\tilde{w}^{\beta*}_{n_2\mathbf{k}}\rangle
=\int
e^{-i\mathbf{q}\cdot\textbf{r}}\tilde{w}^{\alpha}_{n_1\mathbf{k}+\mathbf{q}}(\textbf{r})
\tilde{w}^{\beta*}_{n_2\mathbf{k}}(\textbf{r})\:d^3r\:.
\end{equation}
The tilde denotes the orthonormalized products of Wannier
functions. Although the Wannier functions themselves form an
orthonormal  basis set with $\langle
w^{\alpha}_{n\mathbf{k}}|w^{\beta}_{n^{\prime}\mathbf{k}^{\prime}}\rangle
=N\delta_{nn^{\prime}}\delta_{\mathbf{k}\mathbf{k}^{\prime}}\delta_{\alpha\beta}$,
their  products  do not satisfy this orthonormality condition.
Therefore, we explicitly orthonormalize the products according to
\begin{eqnarray}
\lefteqn{|\tilde{w}^{\alpha}_{n_1\mathbf{k}+\mathbf{q}}
\tilde{w}^{\beta*}_{n_2\mathbf{k}}\rangle} \nonumber \\
&=&
\sum_{n_3,n_4}
\big[O(\mathbf{q},\mathbf{k})^{-1/2}\big]_{n_1n_2,n_3n_4}^{\alpha\beta}
|w^{\alpha}_{n_3\mathbf{k}+\mathbf{q}}w^{\beta*}_{n_4\mathbf{k}}\rangle\:,
\end{eqnarray}
where  the overlap matrix is defined as
\begin{equation}
O_{n_1n_2,n_3n_4}^{\alpha\beta}(\mathbf{q},\mathbf{k})=
\langle w^{\alpha}_{n_1\mathbf{k}+\mathbf{q}}
w^{\beta*}_{n_2\mathbf{k}}
|w^{\alpha}_{n_3\mathbf{k}+\mathbf{q}}
w^{\beta*}_{n_4\mathbf{k}}\rangle\:. \label{overlap}
\end{equation}
In practice, this orthonormalization can be performed in the
final step of the calculation, i.e., in the projection of the
magnetic response function $R$ onto plane waves.

The spin-wave spectra are obtained from the imaginary part of the
transverse magnetic response function $R^{-+}(\mathbf{q},\omega)$,
which exhibits peaks at the spin-wave energies corresponding to
the wave vector $\mathbf{q}$. The half-width of a peak is
inversely proportional to the life time of the excitation.

\subsection{Computational details}

All ground-state calculations are carried out using the FLAPW
method as implemented in the \texttt{FLEUR} code,\cite{Fleur}
initially within the LSDA for the exchange-correlation
potential.\cite{LSDA} We use 4.5 ${\textrm{bohr}}^{-1}$ as a
cutoff for the plane waves  and $l_{\textmd{cut}}=10$ for the
angular momentum for all 3\textit{d} transition metals under
consideration.  In addition, the LSDA+$U$ method with $U=1.9$ eV
and $J=1.2$ eV is employed to reveal the correlation effects on
the magnetic excitation spectra of Ni. In practice, the $U$ and
$J$ values can either be chosen as empirical parameters or
obtained from from first-principles calculations by employing
methods like constrained LSDA\cite{Dederichs:84} or constrained
RPA,\cite{Jepsen,Miyake_U,Miyake} in which the screening due to
the 3\emph{d} electrons is excluded. However, due to \emph{s-d}
hybridization in  the 3\emph{d} transition metals the constrained
RPA yields different values for $U$  and $J$ depending on the
procedure used for excluding 3\emph{d}-3\emph{d} transitions in
the polarization function. For example, Miyake \emph{et
al.}\cite{Miyake_U} found 3.7 eV  for $U$ in fcc Ni if the
\emph{s-d} hybridization is switched off, whereas $U$ reduces to
2.8 eV if the \emph{s-d} hybridization is retained. On the other
hand, our calculations showed that using $U$ and $J$ values from
the constrained LSDA or from constrained RPA within the LSDA+$U$
scheme yields unsatisfactory results for the magnetic moment,
exchange splitting and spin-wave dispersion of fcc Ni compared to
experiments. For this reason we use the empirical values for $U$
and $J$ given above, which improve the Fermi surface and do not
change the magnetic moment substantially.\cite{Savrasov}
Furthermore, these values yield the correct magnetic anisotropy
energy and direction of the magnetization.

The MLWFs are constructed with the \texttt{Wannier90}
code,\cite{Wannier90} which was recently interfaced to the FLAPW
method.\cite{Fleur_Wannier90} The screened Coulomb potential in
the RPA is calculated with the \texttt{SPEX}
code\cite{Friedrich,Friedrich:10} using the mixed product basis
and then projected onto the MLWF basis. The total number of
functions in the mixed product basis is 180-200, and 95-100
unoccupied states are included in the calculation of the
polarizability. Finally, we note that although the MLWFs provide a
minimal basis set for the construction of the $T$-matrix, the
computational time  scales as the  fourth power of the number of
Wannier functions. The most expensive part in our scheme is the
calculation of the kernel $K$,  because it  requires a large
number of $\mathbf{k}$ points for proper convergence (see
Sect.\,\ref{sectionIV}). In contrast, the screened Coulomb
potential $W$ is less sensitive to the $\mathbf{k}$-point
sampling, i.e., it is already converged for a substantially
smaller number of $\mathbf{k}$ points.

\section{Matrix elements of the Coulomb potential}
\label{sectionIII}

\begin{figure}[t]
\begin{center}
\includegraphics[scale=0.28]{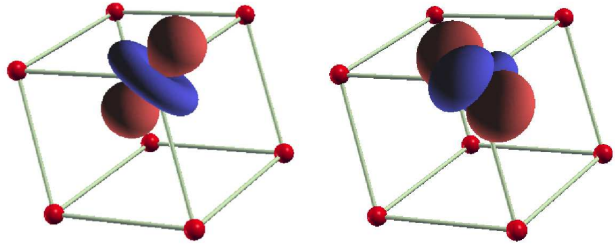}
\end{center}
\vspace*{-0.4cm} \caption{(Color online) $e_{\mathrm{g}}$-like
($3d_{3z^2-r^2}$ and $3d_{x^2-y^2}$) maximally localized Wannier
functions for bcc Fe. The different tints denote regions with
opposite sign.} \label{MLWF_Fe}
\end{figure}

As a first step we calculate the matrix elements of the screened
Coulomb potential $W$ for the series of 3\textit{d} transition
metals, because these matrix elements are a crucial ingredient for
the construction of the magnetic response function. In previous
treatments of spin waves in a  many-body context the Coulomb
interaction was chosen either as a simple Hubbard-type $U$
parameter or as a model potential with an adjustable range
parameter.\cite{Cooke_2,Cooke_3,Karlsson_1} Recent \emph{ab
initio} studies of the bare and the screened Coulomb interaction
in 3\textit{d} transition metals focused only on the nonmagnetic
(NM)
state.\cite{Jepsen,Miyake_U,Springer,Kotani,Schnell,Cococcioni,Solovyev_1,Nakamura}
Here we present a detailed study of the matrix elements in the
MLWF basis for the proper ferromagnetic (FM) state of the
3\textit{d} transition metals Fe, Co, and Ni. For comparison with
previous works the NM states of these three elements and the rest
of the 3\textit{d} series are considered, too. We focus especially
on bcc Fe to investigate the effect of the exchange splitting on
the Coulomb matrix elements, because bcc Fe has the largest
exchange splitting among the 3\emph{d} ferromagnets. Previous
studies showed that, similar to insulators, in 3\emph{d}
transition metals the Wannier functions with \emph{d} character
are exponentially localized.\cite{Schnell} The correlation effects
hence take place predominantly within the same atomic
site.\cite{Miyake} Our calculations confirm these findings. As the
off-site matrix elements of the screened Coulomb potential are at
least two orders of magnitude smaller than the on-site ones, we
only consider the latter. The strong localization of the
3\textit{d} orbitals can be seen in Fig.~\ref{MLWF_Fe}, where we
present $e_{\mathrm{g}}$-like ($3d_{3z^2-r^2}$ and $3d_{x^2-y^2}$)
MLWFs for bcc Fe. The isosurface corresponds to 10\% of the
maximum amplitude of the MLWFs.

To begin with, we define the average on-site diagonal and
off-diagonal matrix elements of the direct and exchange Coulomb
potential as
\begin{equation}
\tilde{U}=  \frac{1}{5}\sum_{n}^{(3d)}W_{nn;nn}^{\alpha\beta}\:,
\end{equation}
\begin{equation}
\tilde{U}^{\prime}= \frac{1}{20}\sum_{\scriptsize{\begin{array}{c} m,n \\
(m\neq n) \end{array}}}^{(3d)}W_{mm;nn}^{\alpha\beta}\:, \\
\end{equation}
\begin{equation}
\tilde{J}= \frac{1}{20}\sum_{\scriptsize{\begin{array}{c} m,n \\
(m\neq n)\end{array}}}^{(3d)}W_{mn;nm}^{\alpha\beta}\:.
\end{equation}
Although the matrix elements of the Coulomb potential are formally
spin-dependent due to the spin-dependence of the  MLWFs, we find
that this dependence is negligible in practice, i.e.,
$W^{\uparrow\uparrow}\simeq W^{\downarrow\downarrow} \simeq
W^{\uparrow \downarrow}$.

\begin{figure}[t]
\begin{center}
\includegraphics[scale=0.58]{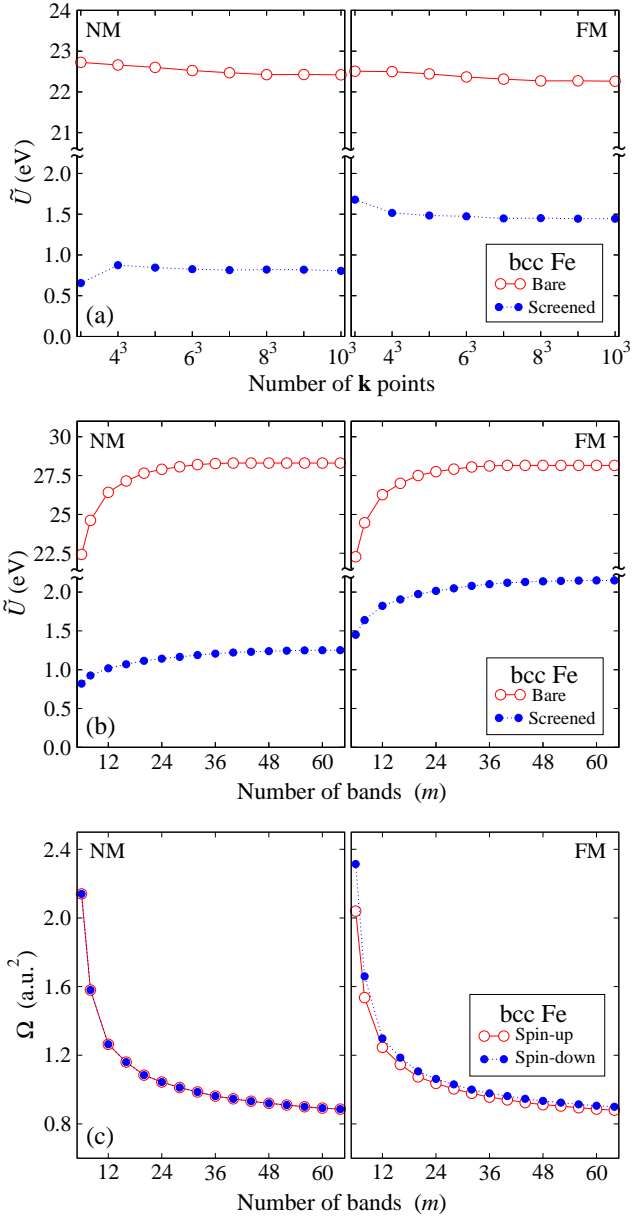}
\end{center}
\vspace*{-0.4cm} \caption{(Color online) Convergence of the
average on-site matrix elements of the bare and screened Coulomb
interaction between the 3\textit{d} orbitals  as a function of (a)
the number of $\mathbf{k}$ points and (b) the number of bands used
in the construction of the MLWF basis for the NM  and FM states of
bcc Fe. (c) The same as (b) for the average spread of the
3\emph{d} orbitals.} \label{convergence}
\end{figure}

In Fig.\,\ref{convergence}(a) we present a convergence study for
the average on-site  diagonal matrix elements  of the bare and the
screened Coulomb interaction ($\tilde{U}$) between the 3\textit{d}
orbitals as a function of the number of $\mathbf{k}$ points for
the NM and the FM states of bcc Fe. Fig.\,\ref{convergence}(b)
shows the same matrix elements as a function of the number of
bands used in the construction of the MLWFs. We observe a fast
$\mathbf{k}$-point convergence but a relatively slow convergence
with respect to the number of bands. In fact, the screened
$\tilde{U}$ is not completely converged even with 60 bands. The
large difference between the NM and FM states will be discussed
below.

The increase of the matrix elements of the Coulomb potential with
the number of bands can be explained by the localization of the
Wannier functions. In Fig.\,\ref{convergence}(c) we show the
spread $\Omega$ for the 3\emph{d} orbitals as a function of the
number of bands. It can clearly be seen that $\Omega$ decreases if
the number of  bands increases, indicating that the 3\emph{d}
orbitals become more localized, which in turn gives rise to a
larger $\tilde{U}$. For the rest of the 3\emph{d} transition-metal
series the behavior of the Coulomb matrix elements with respect to
the number of bands is very similar to bcc Fe. In the rest of this
section  we use 6 bands and an $8\times8\times8$
$\mathbf{k}$-point mesh  in order to compare our results with
previously published data that adopted the same parameter
settings.

In Table\,\ref{table_screen} we present the on-site matrix
elements of the screened Coulomb potential for the NM and FM
states of bcc Fe. Tables\,\ref{table_Co} and \ref{table_Ni_U}
contain the values for the FM state of fcc Co and Ni. For all
three systems the average values for the diagonal ($\tilde{U}$)
and off-diagonal ($\tilde{U}^{\prime}$, $\tilde{J}$) matrix
elements are given in Table\,\ref{table_onsite}. We note that the
splitting of the diagonal matrix elements by the crystal-field
effect is quite pronounced. For bcc Fe the $e_{\mathrm{g}}$-like
diagonal elements are larger than the $t_{2\mathrm{g}}$-like ones
in the FM state, while in the NM state it is just the opposite.
This means that the strongest interaction takes place between the
electrons in the $e_{\mathrm{g}}$-like ($t_{2\mathrm{g}}$-like)
orbitals for the FM (NM) state, because these are more localized.
In the case of fcc Ni the situation is very similar. However, in
fcc Co the splitting of the diagonal Coulomb matrix elements by
the crystal-field effect is different: In the NM state (results
not shown) all diagonal elements assume similar values, while in
the FM state (see Table\,\ref{table_Co}) the
$t_{2\mathrm{g}}$-like diagonal elements are larger than the
$e_{\mathrm{g}}$-like ones.

\begin{table*}[t]
\begin{center}
\caption{Screened on-site direct
($\tilde{U}_{mn}=W_{mm,nn}^{\alpha\beta}$) and exchange
($\tilde{J}_{mn}=W_{mn,nm}^{\alpha\beta}$) Coulomb matrix elements
between the 3\textit{d} orbitals for FM bcc Fe within LSDA. In
parentheses we show results for the NM state. We include 6 bands
in the construction of the MLWFs. The indices 1 and 2 (3, 4, and
5) correspond to the $e_{\mathrm{g}}$-like
($t_{2\mathrm{g}}$-like) Wannier orbitals. All energies are in
eV.}
\begin{ruledtabular}
\begin{tabular}{lccccc}
$\tilde{U}_{mn}$& 1 & 2 & 3& 4& 5 \\ \hline
1 & 1.63 (0.62) &0.35 (0.05)  & 0.64 (0.20) &0.64 (0.20) &0.31 (0.07)  \\
2 & 0.35 (0.05) &1.63 (0.62)  & 0.42 (0.12) &0.42 (0.12) &0.75 (0.25)  \\
3 & 0.64 (0.20) &0.42 (0.12)  & 1.33 (0.96) &0.38 (0.17) &0.38 (0.17)  \\
4 & 0.64 (0.20) &0.42 (0.12)  & 0.38 (0.17) &1.33 (0.96) &0.38 (0.17)  \\
5 & 0.31 (0.07) &0.75 (0.25)  & 0.38 (0.17) &0.38 (0.17) &1.33 (0.96)  \\
\\
$\tilde{J}_{mn}$  & 1 & 2 & 3& 4& 5 \\ \hline
1 & ---           &0.64 (0.28)  & 0.41 (0.30) &0.41 (0.30)  & 0.56 (0.37)\\
4 & 0.64 (0.28) &---            & 0.51 (0.35) &0.51 (0.35   & 0.35 (0.27)\\
3 & 0.41 (0.30) &0.51 (0.35)  & ---           &0.47 (0.41)  & 0.47 (0.41)\\
2 & 0.41 (0.30) & 0.51 (0.35) & 0.47 (0.41) &---           & 0.47 (0.41)\\
5 & 0.56 (0.37) &0.35 (0.27)  & 0.47 (0.41) &0.47 (0.41)  &     ---     \\
\end{tabular}
\end{ruledtabular}
\label{table_screen}
\end{center}
\end{table*}

\begin{table}
\begin{center}
\caption{The same as Table\,\ref{table_screen} for the screened
Coulomb potential for the FM state of fcc Co.}
\begin{ruledtabular}
\begin{tabular}{lccccc}
$\tilde{U}_{mn}$& 1 & 2 & 3& 4& 5 \\ \hline
1&  1.20 &  0.22 &   0.54 &  0.51 &  0.24  \\
2&  0.22 &  1.20 &   0.36 &  0.32 &  0.59  \\
3&  0.54 &  0.36 &   1.45 &  0.38 &  0.39  \\
4&  0.51 &  0.32 &   0.38 &  1.45 &  0.36  \\
5&  0.24 &  0.59 &   0.39 &  0.36 &  1.45  \\
\\
$\tilde{J}_{mn}$& 1 & 2 & 3& 4& 5 \\ \hline
1&  --- &  0.50 & 0.42 & 0.41 &  0.56  \\
2& 0.50 &   ---   & 0.53 & 0.50 &  0.36  \\
3& 0.42 &  0.53 &  ---   & 0.54 &  0.53  \\
4& 0.41 &  0.50 & 0.54 &  ---   &  0.55  \\
5& 0.56 &  0.36 & 0.53 & 0.55 &   ---    \\
\end{tabular}
\end{ruledtabular}
\label{table_Co}
\end{center}
\end{table}

\begin{table}
\begin{center}
\caption{The same as Table\,\ref{table_Co} for fcc Ni.}
\begin{ruledtabular}
\begin{tabular}{lccccc}
$\tilde{U}_{mn}$& 1 & 2 & 3& 4& 5  \\ \hline
1   & 1.53 &  0.28 & 0.57 & 0.58 &  0.26 \\
2   & 0.28 &  1.53 & 0.35 & 0.37 &  0.66 \\
3   & 0.57 &  0.35 & 1.33 & 0.34 &  0.33 \\
4   & 0.58 &  0.37 & 0.34 & 1.33 &  0.33 \\
5   & 0.26 &  0.66 & 0.33 & 0.33 &  1.33 \\
\\
$\tilde{J}_{mn}$& 1 & 2 & 3& 4& 5 \\ \hline
1&   ---   & 0.63 & 0.43 & 0.43 & 0.58 \\
2&  0.63 &  ---   & 0.52 & 0.53 & 0.37 \\
3&  0.43 & 0.52 &  ---   & 0.49 & 0.49 \\
4&  0.43 & 0.53 & 0.49 &  ---   & 0.49 \\
5&  0.58 & 0.37 & 0.49 & 0.49 &  --- \\
\end{tabular}
\end{ruledtabular}
\label{table_Ni_U}
\end{center}
\end{table}

\begin{table}
\caption{Average screened on-site direct (diagonal $\tilde{U}$ and
off-diagonal $\tilde{U}^{\prime}$) and exchange ($\tilde{J}$)
Coulomb matrix elements between the 3\textit{d} orbitals for FM
3\textit{d} transition metals within LSDA. In parentheses we show
results for the NM states. For comparison the LSDA+$U$ results
($U=1.9$ eV, $J=1.2$ eV) for fcc Ni are presented. We include 6
bands in the construction of the MLWFs. All energies are in eV.}
\begin{ruledtabular}
\begin{tabular}{llll}
& $\tilde{U}$    & $\tilde{U}^{\prime}$ &  $\tilde{J}$ \\ \hline
bcc Fe   &    1.45 (0.82) & 0.47 (0.15)  &  0.48 (0.34)\\
fcc Co   &    1.35 (1.04) & 0.39 (0.23)  &  0.49 (0.40)\\
fcc Ni   &    1.41 (1.28) & 0.41 (0.34)  &  0.50 (0.46)\\
fcc Ni$^{*}$  &    1.49        & 0.44         &  0.51 \\
\end{tabular}
\end{ruledtabular}
\label{table_onsite}
\begin{flushleft}
$^{*}$ LSDA+$U$  \\
\end{flushleft}
\end{table}

\begin{figure}[t]
\begin{center}
\includegraphics[scale=0.59]{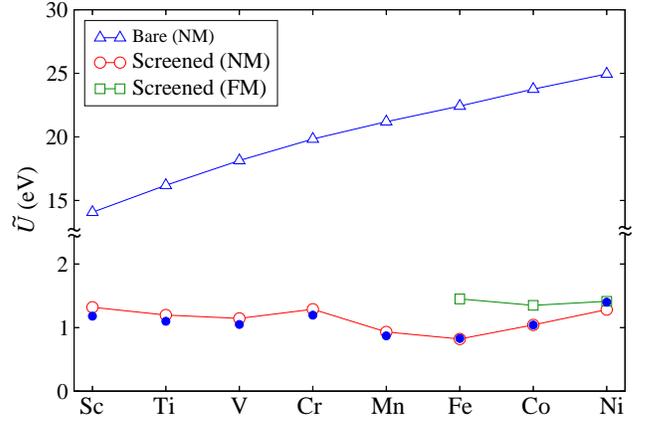}
\end{center}
\vspace*{-0.4cm} \caption{(Color online) Average bare and screened
on-site direct ($\tilde{U}$) Coulomb matrix elements between the
3\textit{d} orbitals for the series of 3\textit{d} transition
metals. For comparison results from Ref.\,\onlinecite{Miyake}
(filled spheres) are given. We include 6 bands in the construction
of the MLWFs.} \label{3d_bare_screen}
\end{figure}

\begin{figure}[t]
\begin{center}
\includegraphics[scale=0.59]{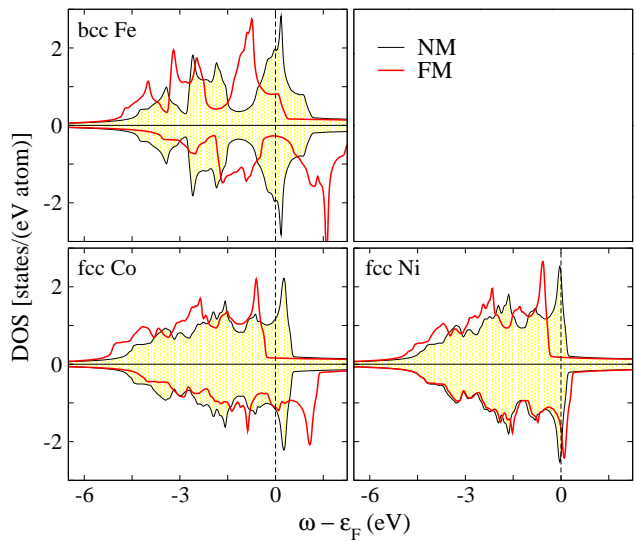}
\end{center}
\vspace*{-0.6cm} \caption{(Color online) Spin-resolved density of
states for bcc Fe, fcc Co, and fcc Ni for the NM and FM states.
The vertical dashed lines denote the Fermi level. Note that
positive values of DOS refer to the majority-spin electrons and
negative values to the minority-spin electrons.} \label{dos_pm_fm}
\end{figure}

Figure\,\ref{3d_bare_screen} shows the average bare and screened
on-site direct ($\tilde{U}$) Coulomb matrix elements for the
series of 3\textit{d} transition metals in the NM state. Results
for the FM state of Fe, Co, and Ni are also included. Note that
among all considered systems only the first three elements are not
magnetic, while Cr and Mn order antiferromagnetically and  Fe, Co,
and Ni are ferromagnetic. The bare $\tilde{U}$ for the NM state
increases linearly from 14 eV for Sc to 25 eV for Ni. This stems
from the fact that, as one moves from the left to the right within
one row of the periodic table, the nuclear charge increases and
causes the 3\textit{d} wave functions to contract. Hence the
localization of the 3\textit{d} electrons increases, giving rise
to the observed trend for $\tilde{U}$. However, this trend is not
observed for the screened Coulomb interaction, where the
calculated values lie between 0.8 eV and 1.5 eV. As  already seen
in Table\,\ref{table_screen}, the matrix elements of the screened
Coulomb potential depend on the magnetic state.
Figure\,\ref{3d_bare_screen} indicates that the $\tilde{U}$ values
for the NM and FM states of Fe, Co, and Ni are also rather
different, and this difference increases with the exchange
splitting in the Ni-Co-Fe sequence. This observation can be
qualitatively explained by the density of states (DOS) around the
Fermi level presented in Fig.\,\ref{dos_pm_fm}. As the screened
Coulomb interaction depends on the polarizability, the number of
occupied and unoccupied states around the Fermi level plays an
important role in determining its strength. Bcc Fe in the NM state
has the largest DOS around the Fermi energy and hence the smallest
Coulomb matrix elements. However, for FM Fe the majority and
minority-spin peaks at the Fermi level are shifted to lower and
higher energies, respectively, due to the exchange field, leading
to a lower DOS at the Fermi level. As a consequence, we obtain
larger matrix elements. For the last three elements  the
calculated $\tilde{U}$, $\tilde{U}^{\prime}$, and $\tilde{J}$
values increase linearly for the NM state (see
Table\,\ref{table_onsite}), while they are almost constant for the
FM state. A comparison of our results for the screened Coulomb
interaction in the NM state with Ref.\,\onlinecite{Miyake} is
given in Fig.\,\ref{3d_bare_screen}. The agreement for the
$\tilde{U}$ values is very good, and we find an equally good
agreement for the $\tilde{J}$ values, which are not displayed in
the figure.

To investigate the effect of static correlation on the matrix
elements of the screened Coulomb interaction  we present LSDA+$U$
results for $\tilde{U}$, $\tilde{U}^{\prime}$, and $\tilde{J}$ for
the FM state of Ni in Table\,\ref{table_onsite}.  The average
matrix elements slightly increase relative to the LSDA. Again this
observation can be explained by the scenario given above. Within
the LSDA+$U$ scheme the exchange splitting of the Ni 3\emph{d}
states increases. This in turn gives rise to a larger magnetic
moment (see Table\,\ref{table_Ni}) and a reduced DOS around the
Fermi level. As a consequence, the Coulomb matrix elements
increase. We expect a similar behavior for bcc Fe and fcc Co if
the LSDA+$U$ scheme is employed.

Finally, we discuss the values of the matrix elements of the
screened Coulomb potential $W$ used in the calculation of $R$ to
make a connection with next section. The magnetic response
function can be schematically written as
\begin{equation}
R=\frac{K}{1-WK}\:,
\label{Schematic_R}
\end{equation}
where the screened Coulomb potential $W$ in the denominator is
responsible for the formation of  collective spin-wave
excitations. As shown above, the matrix elements of $W$ depend on
the number of bands included in the construction of the MLWFs. The
matrix elements of $K$ are considerably less sensitive to the
number of bands; usually 10 bands are sufficient for 3\emph{d}
ferromagnets. For this reason, in the calculation of $R$ one might
choose a particular $W$ that satisfies the exact condition
$\lim_{\mathbf{q}\rightarrow \mathbf{0}} \omega(\mathbf{q})=0$
(Goldstone mode). For instance, in fcc Ni within LSDA one should
include about 100 bands in order to fulfill the Goldstone theorem.
Alternatively, one can calculate $W$ for a given number of bands
and then scale it by a factor $\eta$, i.e., $W \rightarrow \eta
W$, to obtain the Goldstone mode correctly. This second approach
is computationally less demanding and is used in the present work.
For 3\emph{d} ferromagnets we calculate $W$ by including only 6
bands. In the case of fcc Ni the scaling factor is $\eta \approx
1.5$ within LSDA and $\eta \approx 1.8$ within LSDA+$U$. This
means that the Coulomb matrix elements presented in
Table\,\ref{table_Ni_U} should be multiplied by 1.5 in the
calculation of the magnetic response function $R$ within LSDA for
fcc Ni. The violation of the Goldstone theorem  within the present
formalism stems from the approximations made in the calculation of
the kernel $K$ and the screened Coulomb potential $W$. In a fully
self-consistent linear-response calculation without additional
approximations the Goldstone theorem should be fulfilled. However,
this is hardly feasible in practice. When we manually reduce the
exchange splitting of Ni by one-half within LSDA to simulate the
renormalized Green function, the scaling factor is reduced to
$\eta \approx 1.1$. Note that the LSDA overestimates the exchange
splitting of Ni by a factor of 2 compared to
experiments.\cite{exc_1,exc_2} In systems like bcc Fe for which
the LSDA already provides a reasonable description of the
electronic band structure compared to the
experiments\cite{exc_2,Singh,Schindlmayr} the scaling factor
$\eta$ is close to 1.

\section{Magnetic excitations in $\textrm{fcc Ni}$}
\label{sectionIV}

This section deals with magnetic excitations in fcc Ni. Among the
3\textit{d} ferromagnets Ni is known for particularly large
discrepancies between the results from DFT calculations and
experiments: The width of the occupied 3\textit{d} bands in the
LSDA is about 30\% larger than that found in photoemission
experiments, whereas the \textit{sp} band width agrees within
10\%.\cite{band_width,band_width_2} Similarly, the LSDA yields a
much smaller DOS [1.9 states/(eV atom)] at the Fermi level
compared to low-temperature specific-heat data [3.0 states/(eV
atom)], indicating a quasiparticle mass
enhancement.\cite{dos_fermi} Even larger discrepancies are
obtained for the exchange splitting. Photoemission experiments
give a small and highly anisotropic exchange splitting, 0.3 eV at
the $\textrm{L}_3$ point and 0.2 eV at the $\textrm{X}_2$
point.\cite{exc_1,exc_2} In contrast, the LSDA yields a rather
large (0.6 eV) and almost isotropic splitting.\cite{band_width}
However, the calculated magnetic moment turns out to be in good
agreement with the experimental value.\cite{Ni_moment}

\begin{figure}
\begin{center}
\includegraphics[scale=0.58]{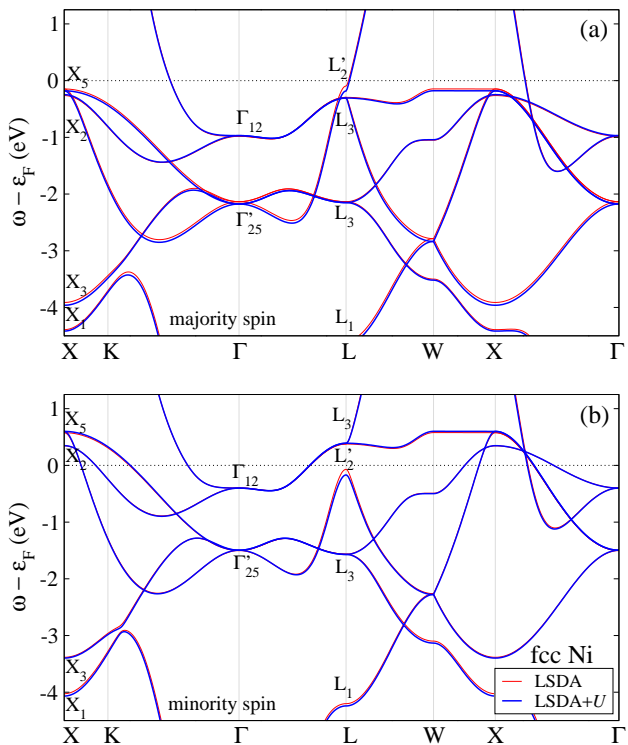}
\end{center}
\vspace*{-0.4cm} \caption{ (Color online) LSDA and LSDA+$U$
($U=1.9$ eV and $J=1.2$ eV) band structures of FM fcc Ni for (a)
majority and (b) minority spins.} \label{band_structure}
\end{figure}

Our calculated  magnetic moments and exchange splittings within
the LSDA and LSDA+$U$ are presented in Table\,\ref{table_Ni}. The
corresponding band structures are given in
Fig.\,\ref{band_structure}. Within the LSDA the band structure,
exchange splitting, and magnetic moment are in good accordance
with literature values.\cite{band_width,band_width_2,Ni_moment}
The inclusion of explicit static correlation in the form of a
Hubbard $U$ within LSDA+$U$ slightly changes the electronic
structure of Ni. In this case the exchange splitting is more
anisotropic compared to the LSDA, i.e., it increases at the
$\mathrm{L}_3$ point and decreases at the $\mathrm{X}_2$ point
(see Table\,\ref{table_Ni}). The average exchange splitting
increases within the LSDA+$U$ scheme however, and depending on the
values of the Hubbard parameters $U$ and $J$ the LSDA+$U$ hence
gives rise to a larger magnetic moment.

\begin{table}
\caption{Calculated magnetic moment $m$,
exchange splitting at  $\textrm{X}_2$ and $\textrm{L}_3$ as well
as spin-wave stiffness constant $D$ in fcc Ni
within LSDA and LSDA+$U$ with $U=1.9$ eV and
$J=1.2$ eV. For comparison experimental values are given.}
\begin{ruledtabular}
\begin{tabular}{lllll}
         & $m$($\mu_{\mathrm{B}}$)  & $\textrm{X}_2$(eV) & $\textrm{L}_3$(eV) & $D(\textrm{meV}\mathrm{{\AA}}^2$) \\
         \hline
LSDA     &  0.61       &  0.61       &  0.57 &   740  \\
LSDA+$U$ &  0.65       &  0.55       &  0.62 &    540 \\
Expt.    &  0.60$^{a}$ &  0.20$^{b}$ &  0.30$^{c}$ & 550$^{d}$\\
\end{tabular}
\label{table_Ni}
\end{ruledtabular}
\begin{flushleft}
$^{a}$Ref.\,[\onlinecite{Ni_moment}]\\
$^{b}$Ref.\,[\onlinecite{exc_1}]\\
$^{c}$Ref.\,[\onlinecite{exc_2}] \\
$^{d}$Ref.\,[\onlinecite{Mook_2}]
\end{flushleft}
\end{table}

The magnetic response function $R$ [see Eq.\,(\ref{Schematic_R})]
contains all relevant information about the dynamics of the spin
system. The poles of $K$ in the numerator correspond to the
energies of single-particle spin-flip Stoner excitations, while
the zeros of the denominator $(1-WK)$ describe collective
spin-wave excitations. In the preceding section we showed  that
the convergence of the matrix elements of the screened Coulomb
potential $W$ with respect to the number of $\mathbf{k}$ points
(number of bands) is fast (slow), whereas the situation is
opposite for the kernel $K$. For $W$ we take the values from
Sec.\,\ref{sectionIII}, scaled by an appropriate factor $\eta$,
while we construct $K$ using 5 MLWFs, 15 occupied and unoccupied
bands per spin channel, and a very dense $40\times40\times40$
$\mathbf{k}$-point sampling. The Brillouin-zone summations in $K$
are performed with the tetrahedron method.\cite{Tetrahedron} The
calculation of $R$ is carried out for a fixed $\mathbf{q}$ as a
function of energy $\omega$ up to 1.5 eV.

The remainder of this section is divided into two parts. In the
first part the single-particle Stoner excitations are presented.
The second part deals with the collective spin-wave excitations.

\subsection{Single-particle Stoner excitations}

Stoner excitations are electron transitions between bands of
opposite spin. When an electron is excited from an occupied
majority-spin state at $\mathbf{k}$ to an unoccupied minority-spin
state at $\mathbf{k}$+$\mathbf{q}$, it produces an electron-hole
pair with triplet spin configuration that reduces the
magnetization by unity. Therefore, these excitations are
associated with longitudinal fluctuations of the magnetization and
play an important role in determining the high-temperature
properties of magnetic materials.\cite{Moriya} To simplify the
discussion, let us consider a free electron gas with parabolic
dispersion and a rigid exchange splitting $\Delta
E_{\textrm{ex}}$. The corresponding single-particle excitation
energies are given by $\omega(\mathbf{q})=(\pm 2\mathbf{k} \cdot
\mathbf{q}+\mathbf{q}^2)+ \Delta E_{\textrm{ex}}$. For
$\mathbf{q}\rightarrow \mathbf{0}$ the necessary energy for such
excitations equals the exchange splitting $\Delta
E_{\textrm{ex}}$, but for finite $\mathbf{q}$ these excitations
form a continuous spectrum, which is called the \emph{Stoner
continuum} and is determined by all possible values of
$\mathbf{k}$. The upper and lower bounds of the Stoner continuum
depend on the electronic band structure of the material. For
example, in strong ferromagnets the smallest possible excitation
energy is given by the Stoner gap $\Delta S$. On the other hand,
weak ferromagnets exhibit a vanishing gap $\Delta S=0$, and thus
single-particle excitations do not require a finite energy for
particular $\mathbf{q}$ values. The lower bound is of special
interest, because when the  collective spin-wave excitations enter
the Stoner continuum, they start to decay into single-particle
excitations, which reduces their lifetime drastically. We note
that the picture of Stoner excitations in real materials is
different  from the free electron gas with a single
band.\cite{Kuzemsky}

\begin{figure}[t]
\begin{center}
\includegraphics[scale=0.58]{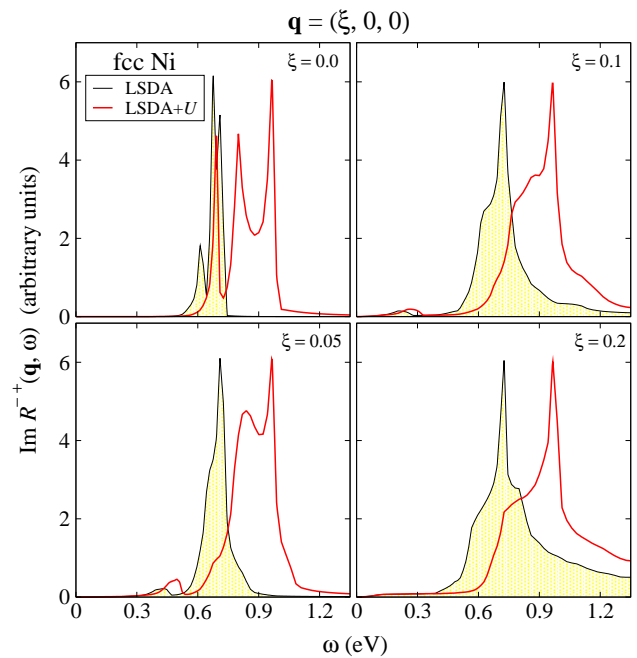}
\end{center}
\vspace*{-0.5cm} \caption{(Color online)  The Kohn-Sham magnetic
response function for fcc Ni for selected wave vectors along the
$\Gamma - \textrm{X}$ direction in the Brillouin zone. Since the
Kohn-Sham system is noninteracting, the spectrum  exhibits only
single-particle Stoner excitations but no collective magnon modes
at lower energies. In each panel the peak amplitudes are scaled to
the same height.} \label{stoner_excitations}
\end{figure}

Since the Stoner excitations are single-particle spin-flip
processes, they can be studied qualitatively at the Kohn-Sham
level. They result from the term $\textrm{Im}\, K$, as discussed
before. In Fig.\,\ref{stoner_excitations} we present the imaginary
part of the Kohn-Sham magnetic response function for fcc Ni for
selected wave vectors along the $\Gamma - \textrm{X}$ direction.
Note that all peak amplitudes are scaled to the same height. The
energetic position and shape of the Stoner spectrum for
$\mathbf{q}=\mathbf{0}$ provide a measure for the mean exchange
splitting and its variation across the Brillouin zone. The
triple-peak structure reflects the intraband and interband
transitions. The position of the first peak gives the mean
exchange splitting: 0.61 eV within LSDA and 0.69 eV within
LSDA+$U$. The broadening of the peaks reflects the
$\mathbf{k}$-dependence of the exchange splitting. For rigidly
split bands one would get $\delta$ peaks. As the wave vector
increases, individual peaks become broader and are no longer
distinguishable. Since the Kohn-Sham system is noninteracting, the
spectrum does not exhibit collective magnon modes at low energies.

Experimentally, Stoner excitations in 3\emph{d} ferromagnets are
studied with spin-polarized electron-energy-loss spectroscopy
(SPEELS), a technique that not only measures the high-energy
Stoner excitations but also low-energy collective spin-wave modes
up to the Brillouin-zone boundary. Using SPEELS Kirschner
\textit{et al.}\cite{Kirschner_1} studied the Stoner excitation
spectrum of Ni at $\mathbf{q}\approx \mathbf{0}$. The authors
found that the spectrum had a broad energy distribution of 0.3 eV
(full width at half maximum) and was centered around 0.3 eV, which
is consistent with the average exchange splitting determined by
photoemission experiments. Additionally, the width of the
distribution provided evidence for the pronounced
$\mathbf{k}$-dependence of the exchange splitting. The comparison
of our calculated spectra with the experimental data shows large
discrepancies, however. These can be attributed to the
overestimated exchange splitting and its incorrect nearly
isotropic behavior over the Brillouin zone in the LSDA. As pointed
out by Oles and Stollhoff\cite{Oles} and by Liebsch,\cite{Liebsch}
the large exchange splitting in the LSDA is due to the neglect of
strong correlation effects within the 3\emph{d} states and
anisotropic exchange. Consequently, recent studies by Katsnelson
and Lichtenstein\cite{DMFT_1} and by Grechnev \emph{et
al.}\cite{DMFT_2} based on dynamical mean-field theory, in which
the exchange and correlation effects are taken into account
properly, gave much improved results for the exchange splitting
and quasiparticle band structure of Ni.

\subsection{Collective spin-wave excitations}

In the preceding section we discussed the non-interacting magnetic
response function $K$, which has singularities that correspond to
single-particle Stoner excitations, usually with a broad energy
distribution. In addition, there may be other singularities of $R$
for a fixed $\mathbf{q}$. For some $\omega(\mathbf{q})$ outside
the Stoner continuum the denominator ($1-WK$) of the magnetic
response function might vanish, indicating collective spin-wave
excitations. These correspond to transverse fluctuations of the
direction of the magnetization and can be interpreted as a
coherent superposition of an electron and a hole with opposite
spins coupled via an attractive screened Coulomb interaction $W$,
thereby forming a bound state with energy $\omega(\mathbf{q})$,
spin $1$, and momentum $\mathbf{q}$.\cite{Pines} As more than one
particle is involved in the excitation process, the formation of
spin waves cannot be described within a single-particle picture.

Before discussing our numerical results we briefly review the
status of experimental and theoretical studies of magnetic
excitations in fcc Ni. Starting from the mid 1960s the spin
dynamics of ferromagnetic 3\textit{d} transition metals and their
alloys were intensively investigated by inelastic neutron
scattering. The first experiments for Ni were performed at high
temperatures and in the low-energy region. Later the measurements
were extended to higher energies. The results of these early
neutron-scattering experiments and a comparison with realistic
band structure calculations can be found in the review article by
Lowde and Windsor.\cite{Windsor} However, the signal intensity and
energy resolution in these early experiments were not favorable
for a quantitative determination of the spin-wave excitations in
Ni. More precise measurements were reported by Mook and
collaborators in the 1980s with emphasis on high energies up to
240 meV.\cite{Mook_1,Mook_2,Mook_3} The authors  measured the
spin-wave dispersion of Ni at several temperatures starting from
$T = 4.2\,\textrm{K}$ up to $T \approx 2T_{\mathrm{C}}$, where
$T_{\mathrm{C}}=631\,\textrm{K}$ is the Curie temperature,  and
found that the spin-wave dispersion was isotropic in $\mathbf{q}$
over the entire temperature range studied. The obtained spin-wave
stiffness constant was $D=550$ meV $\mathrm{{\AA}}^2$ at
$T=4.2\,\textrm{K}$ and $D=505$ meV $\mathrm{{\AA}}^2$ at
$T=295\,\textrm{K}$. The spin-wave intensity was found to decrease
faster with increasing energy in the $[1\,1\,1]$ direction than
along other symmetry directions. In all directions the  spin waves
eventually disappeared at some wave vector close to the zone
boundary inside the first Brillouin zone, which was attributed to
a decay into Stoner excitations. Additionally, the authors found
evidence for a second branch in the spin-wave dispersion, i.e., an
optical mode that crosses the main $[1\,0\,0]$ acoustic branch
around 125 meV. However, inelastic neutron scattering does not
allow to explore the entire Brillouin zone in 3\emph{d}
ferromagnets, in contrast to SPEELS. Using SPEELS Abraham and
Hopster\cite{Abraham} attempted to study short-wave-length (or
large-wave-vector) spin excitations in Ni. However, they only
detected Stoner excitations, although the data reported reaches
down to 100 meV and  the resolution of the instrument (17 meV)
should have been sufficient to observe collective excitations. A
qualitative explanation for the absence of spin-wave peaks in the
SPEELS spectrum was given by Hong and Mills,\cite{Hong} who showed
that the spin waves can only be observed in SPEELS if the exchange
splitting of the 3\emph{d} bands is large compared to the
spin-wave excitation energies.  Indeed, spin waves up to the
Brillouin-zone boundary are observed in Fe and Co, which have
substantially larger exchange splittings than
Ni.\cite{Speels_1,Speels_2,Speels_3,Speels_4}

On the theoretical side, the first attempt to use realistic energy
bands for Ni in the calculations of a generalized susceptibility
was undertaken by Lowde and Windsor.\cite{Windsor} The authors
calculated the magnetic susceptibility within the RPA for rigidly
spin-split bands, but the agreement with the available
experimental data was not good.  Cooke \textit{et
al.}\cite{Cooke_1,Cooke_2,Cooke_3} showed that it is necessary to
take the $\mathbf{k}$-dependence of the exchange splitting into
account for a quantitative comparison between theory and
experiment. They used a tight-binding description of the
electronic energy bands, and the ferromagnetism was driven by an
empirical on-site Coulomb interaction between the 3\emph{d}
electrons with two adjustable parameters. These parameters were
chosen in such a way that the calculations reproduce the
experimentally observed magnetic moment as well as the correct
$t_{2\mathrm{g}}$ and $e_{\mathrm{g}}$ character of the moment as
measured in neutron magnetic-form-factor experiments. The
calculations of Cooke \textit{et al.} not only yielded the correct
spin-wave dispersion relation, including the appearance of the
optical branch in the $[1\,0\,0]$ direction, but the damping of
the spin waves in the presence of the Stoner modes was also
correctly described. A similar approach was used by Hong and
Mills\cite{Hong} with an empirical Coulomb interaction that was
form-invariant under spin rotations. However, they failed to find
the optical mode in the spin-wave dispersion of Ni.

A much more accurate description of spin waves in ferromagnetic
3\emph{d} transition metals was reported by
Savrasov\cite{Savrasov_SW} and by Karlsson and
Aryasetiawan.\cite{Karlsson_1} In both works spin-polarized DFT
was used for the ground-state calculations. For the transverse
spin susceptibility Savrasov employed TDDFT, while Karlsson and
Aryasetiawan adopted MBPT. Similar to the findings of Cooke
\textit{et al.}, Savrasov obtained two branches in the spin-wave
dispersion of Ni along the $[1\,0\,0]$ direction with an optical
mode at high energies. Karlsson and Aryasetiawan confirmed these
results  and investigated the role of the one-particle band
structure on the spin-wave dispersion. They found a very good
agreement between theory and experiment if the LSDA exchange
splitting was manually reduced by one-half. Additionally, the
calculations gave evidence for an optical branch along the
$[1\,1\,1]$ direction.

\begin{figure}[t]
\begin{center}
\includegraphics[scale=0.58]{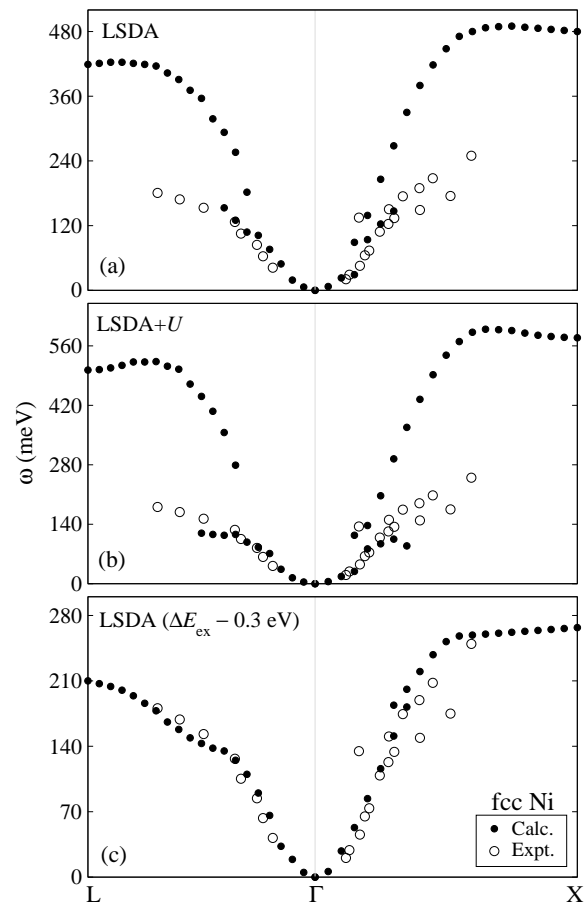}
\end{center}
\vspace*{-0.5cm} \caption{Spin-wave dispersion for fcc Ni along
high-symmetry lines $(\textrm{L}-\Gamma-\textrm{X})$ in the
Brillouin zone within (a) LSDA, (b) LSDA+$U$, and (c) LSDA with a
reduced exchange splitting. The experimental dispersion is taken
from Refs.\,\onlinecite{Mook_2} and \onlinecite{Mook_3}.}
\label{spin_wave_dispersion}
\end{figure}

The optical branch in the spin-wave dispersion of Ni is evidently
a very subtle issue. So far, there has been no general consensus
in theoretical treatments concerning the sensitivity of the
results to the details of the electronic structure and to the
method used. In the following we hence focus on the energy region
where the double-peak structure is observed in the spin-wave
spectrum. To illuminate the effect of the electronic structure on
the magnetic excitation spectrum of Ni we employ three different
methods: LSDA, LSDA+$U$, and LSDA with a reduced exchange
splitting by one-half. The calculated spin-wave dispersion along
the high-symmetry lines $\textrm{L}-\Gamma-\textrm{X}$ is
displayed in Fig.\,\ref{spin_wave_dispersion}. For comparison the
experimental dispersion is also shown. The LSDA and LSDA+$U$ yield
qualitatively similar results, with an optical mode not only in
the $[1\,0\,0]$ but also in the $[1\,1\,1]$ direction. The
acoustic branch is well described within LSDA and LSDA+$U$, but
the optical branch is too high in energy. This discrepancy between
theory and experiment can be traced back to the overestimation of
the exchange splitting in the LSDA.\cite{Karlsson_1} Indeed, when
we reduce the exchange splitting by one-half as in
Ref.\,\onlinecite{Karlsson_1}, we obtain reasonable agreement with
the experiments. The corresponding dispersion, shown in
Fig.\,\ref{spin_wave_dispersion}(c), is similar to that of
Karlsson and Aryasetiawan.\cite{Karlsson_1}

\begin{figure}[t]
\begin{center}
\includegraphics[scale=0.58]{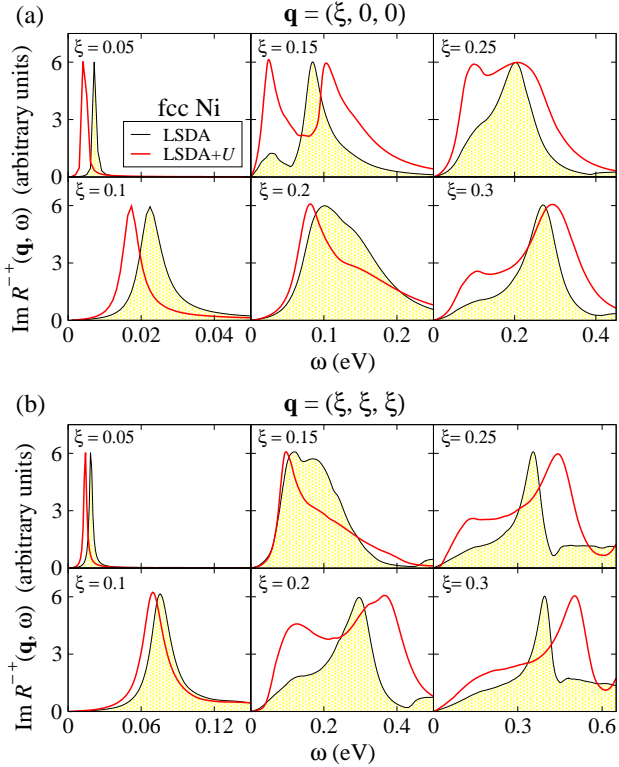}
\end{center}
\vspace*{-0.5cm} \caption{(Color online) Spin-wave excitation
spectra for fcc Ni within LSDA and LSDA+$U$ for selected wave
vectors along (a) the  $[1\,0\,0]$ and (b) the $[1\,1\,1]$
directions in the Brillouin zone. The double-peak structure for
certain wave vectors is clearly visible. All peak amplitudes are
scaled to the same height.} \label{magnon_spectra}
\end{figure}

\begin{figure}[t]
\begin{center}
\includegraphics[scale=0.56]{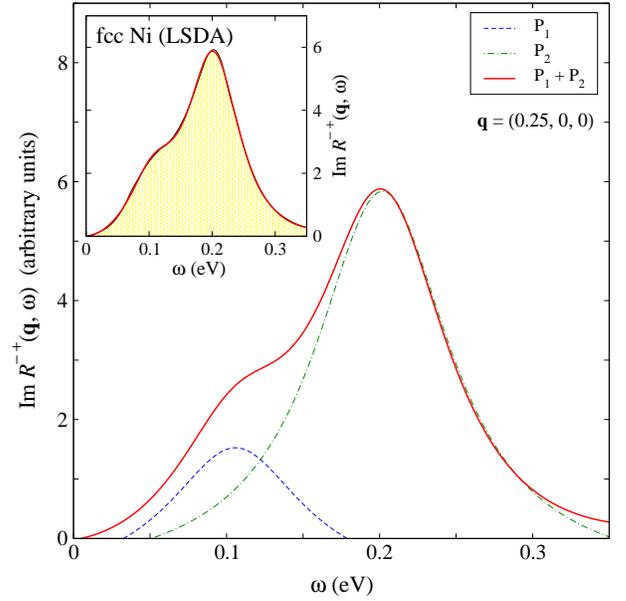}
\end{center}
\vspace*{-0.5cm} \caption{(Color online) Decomposition of the
single feature for $\mathbf{q}=(0.25, 0, 0)$ into two Lorentzian
peaks. In the inset  we compare with the original curve.}
\label{double_peaks}
\end{figure}

In Fig.\,\ref{magnon_spectra} we show the imaginary part of the
magnetic response function $R$ for selected wave vectors along the
$[1\,0\,0]$ and  $[1\,1\,1]$ directions. The peak amplitudes are
again scaled to the same height for presentational purposes. The
obtained spin-wave spectra along $[1\,0\,0]$ are in very good
agreement with previous calculations.\cite{Savrasov_SW,Karlsson_1}
In particular, a double-peak structure starts to develop from
$\mathbf{q}=(0.15,\,0,\, 0)$. For larger wave vectors the two
peaks overlap, resulting in a rather broad single feature, which
can be decomposed into two Lorentzian peaks as shown in
Fig.\,\ref{double_peaks} for $\mathbf{q}=(0.25,\,0,\,0)$. As the
wave vector increases, the intensity of the lower peak decreases
in agreement with experiments.\cite{Mook_3} In the $[1\,1\,1]$
direction the double peak is not so clear within the LSDA.
Nevertheless, the calculated structure can still be decomposed
into two Lorentzians (results not shown), but the $\mathbf{q}$
interval where the double-peak structure appears is small compared
to the $[1\,0\,0]$ direction.

\begin{figure}[t]
\begin{center}
\includegraphics[scale=0.58]{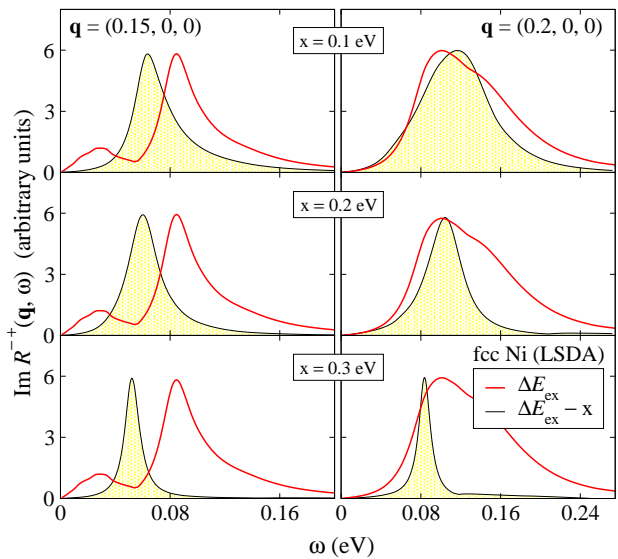}
\end{center}
\vspace*{-0.5cm} \caption{(Color online) Same as
Fig.\,\ref{magnon_spectra} for $\mathbf{q}=(0.15,0,0)$ and
$\mathbf{q}=(0.2,0,0)$  for different exchange splitting within
LSDA. The spectra for the original LSDA exchange splitting are
included for comparison.} \label{exchange_splitting}
\end{figure}

Our calculations show that the exchange splitting has a strong
influence on the emergence of a double-peak structure. To
demonstrate this we consider two $\mathbf{q}$ points in the
$[1\,0\,0]$ direction where such features appear and calculate the
LSDA spin-wave spectra by reducing the exchange splitting
gradually by 0.1 eV in each step, up to one-half of its original
value. The results are presented in
Fig.\,\ref{exchange_splitting}. Even a very small reduction in the
exchange splitting by 0.1 eV strongly suppresses the double-peak
feature. As the exchange splitting reduces further, the width of
the peaks becomes narrower, indicating an increased spin-wave life
time. When the exchange splitting is reduced by one-half and
corresponds to the experimental value, the optical branch
completely disappears in the $[1\,1\,1]$ direction, while it is
shifted to larger wave vectors in the $[1\,0\,0]$ direction. The
situation is very similar within LSDA+$U$ with a reduced exchange
splitting. Karlsson and Aryasetiawan found a weak double-peak
structure with a reduced exchange splitting for
$\mathbf{q}=0.1875(1,\,1,\,1)$ when they used a small Gaussian
broadening parameter in the Brillouin-zone
integration,\cite{Karlsson_1} but this double-peak structure
becomes smeared out if a larger broadening is used. Based on the
symmetry properties of fcc Ni, the spin-wave dispersion should be
isotropic, i.e., one expects an optical branch also in the
$[1\,1\,1]$ direction, but the resolution of the presently
available experimental data is not sufficient to observe
it.\cite{Mook_3} Of course, a reduction of the exchange splitting
in the calculations does not solve all problems for Ni. In
particular, too large band widths and the incorrect isotropic
exchange splitting remain important issues whose effect on the
spin-wave spectra requires further investigations.

As pointed out by Cooke \emph{et al.},\cite{Cooke_2} the
double-peak structure in the magnetic excitation spectra of
3\emph{d} ferromagnets stems from the $\mathbf{k}$-dependent
exchange splitting and interband transitions. A detailed analysis
was given by Karlsson and Aryasetiawan,\cite{Karlsson_1} who
showed that the double-peak structure in fcc Ni is implicitly
contained in the kernel $K$, i.e., it is a band-structure effect
arising from the fact that $W\textrm{Im}\,K$ possesses  additional
structure below the Stoner peak, which in turn gives rise to a
weak dip structure in $(1-W\textrm{Re}\,K)$ via the Hilbert
transformation. Such a weak dip structure appears as a second peak
in the spin-wave excitation spectrum for particular wave vectors.
This observation is indirectly confirmed by our analysis. The
application of a Hubbard $U$ in the LSDA+$U$ calculation shifts
majority-spin and minority-spin states more or less rigidly, and
the optical branch appears in the LSDA and LSDA+$U$ around the
same wave vector $\mathbf{q}$ along $[1\,0\,0]$, while a reduction
of the exchange splitting leads to an appearance of the optical
branch at higher $\mathbf{q}$ values.

The magnetic response function $R$ allows to extract information
about the spin-wave life times, which are inversely proportional
to the widths of the spin-wave peaks. If Ni were a strong
ferromagnet, one would get well defined $\delta$ peaks up to about
0.3 eV in the spin-wave spectra, corresponding to the energy
difference between the highest occupied majority-spin 3\textit{d}
band and the Fermi level, i.e., the Stoner gap. However, Ni is not
a truly strong ferromagnet due to the \emph{sp-d} hybridized
majority states around the Fermi energy, and Stoner excitations
thus occur essentially at all energies. This means that spin waves
decay into Stoner excitations for all non-zero wave vectors, which
is reflected by a finite width of the spin-wave peaks as shown in
Fig.\,\ref{magnon_spectra}. The life time of the spin waves
depends on the details of the coupling between these excitations
and on the density of states of the Stoner excitations. It should
be noted that in itinerant ferromagnets the life time of the spin
waves becomes infinite in the limit of a vanishing wave vector,
provided that spin-orbit coupling is ignored. As seen in
Fig.\,\ref{magnon_spectra}, for small wave vectors the spin-wave
peaks are indeed narrow, and the damping is hence weak, i.e., the
life times are long. As the wave vector increases, the spin-wave
dispersion enters into the region with a high density of states of
Stoner excitations, and the decay mechanism becomes more
efficient. In both crystallographic directions considered here the
maximum damping occurs in the region where the double-peak
structure appears. In contrast to the findings of Cooke \emph{et
al.,}\cite{Cooke_2} the spin-wave peaks associated with the
optical branch are narrower than the acoustic ones  in our work,
and the peak widths stay almost constant throughout the Brillouin
zone, while the peak widths in the acoustic branch increase with
the wave vector. This might in fact explain why the acoustic
branch disappears in the middle of the Brillouin zone in the
neutron-scattering experiments.\cite{Mook_3}

Finally, we focus on the spin-wave stiffness constant $D$. In
ferromagnets  the spin waves show a quadratic dispersion law
$\omega(\mathbf{q})=Dq^2$ for small wave vectors. The values for
Ni obtained with different methods are listed in
Table\,\ref{table_Ni}. Our LSDA estimate of
$740\,\textrm{meV}\,\mathrm{{\AA}}^2$ is substantially larger than
the experimental value $550\,\textrm{meV}\,\mathrm{{\AA}}^2$. With
the reduced exchange splitting $D$ increases even further to
$870\,\textrm{meV}\,\mathrm{{\AA}}^2$, whereas LSDA+$U$ provides a
much better estimate of $540\,\textrm{meV}\,\mathrm{{\AA}}^2$,
which reflects the importance of static correlation effects in Ni.
We note that our LSDA estimate for the spin-wave stiffness of fcc
Ni is in good agreement with calculations based on constrained DFT
in the adiabatic approximation. Using the frozen-magnon technique
Rosengaard and Johansson\cite{Johansson} found
$D=739\,\textrm{meV}\,\textrm{{\AA}}^2$, which is very similar to
the values obtained by Schilfgaarde and
Antropov\cite{Schilfgaarde}
($740\,\textrm{meV}\,\textrm{{\AA}}^2$) and by Pajda \textit{et
al.}\cite{Pajda} ($756 \pm 29\,\textrm{meV}\,\textrm{{\AA}}^2$),
who employed real-space methods. As the adiabatic approximation
becomes exact in the limit of long wave lengths ($\mathbf{q}
\rightarrow \mathbf{0}$),\cite{Katsnelson_1} this can be compared
with values obtained from more rigorous approaches based on the
dynamical transverse spin susceptibility or magnetic response
function. However, within constrained DFT the second branch in the
spin-wave dispersion of Ni cannot be described.

\section{Conclusions and outlook}
\label{sectionV}

In conclusion, we have developed a computational method to study
excitation spectra of magnetic materials from first principles.
The method is based on many-body perturbation theory. The main
quantity of interest is the transverse magnetic response function,
which treats both collective spin-wave excitations and
single-particle spin-flip Stoner excitations on an equal footing.
In order to describe the former we include appropriate vertex
corrections in the form of a multiple-scattering $T$-matrix, which
describes the coupling of electrons and holes with different
spins. To reduce the numerical cost for the calculation of the
four-point $T$-matrix we exploit a transformation to maximally
localized Wannier functions that takes advantage of the short
spatial range of electronic correlation in the partially filled
\textit{d} or \textit{f} orbitals of magnetic materials. Our
implementation is based on the FLAPW method.

The developed scheme was employed to calculate the matrix
elements of the Coulomb potential for the series of 3\emph{d}
transition metals in the MLWF basis. Special attention was given
to the ferromagnets Fe, Co and Ni. We showed that the matrix
elements of the screened Coulomb potential are rather different
for the NM and FM states and that the difference increases with
the exchange splitting in the Ni-Co-Fe sequence, which can be
accounted for on the basis of the total density of states around
the Fermi level for the corresponding systems.

The magnetic excitations in fcc Ni were studied in detail based
on the LSDA and LSDA+$U$ methods. Both schemes give qualitatively
similar results for the spin-wave spectra and dispersion. However,
correlation effects seem to be important for the spin-wave
stiffness constant, which is overestimated within LSDA. Our
calculations indicate the existence of an optical branch in the
spin-wave dispersion of Ni along the $[1\,1\,1]$ in addition to
that along the $[1\,0\,0]$ direction in the Brillouin zone.
Although the acoustic branch is well described within LSDA and
LSDA+$U$, the optical branch appears to be too high in energy.
This discrepancy between theory and experiment can be attributed
to the overestimation of the exchange splitting or, in other
words, to the use of the Kohn-Sham Green function in the
calculation of the kernel $K$ instead of the renormalized one.

In the LSDA Kohn-Sham Green function the long and short-range
correlation effects are not taken into account properly. The
former can be treated within the  $GW$ approximation, while the
latter require the summation of spin-dependent $T$-matrix
contributions. In the future we plan to incorporate
electron-electron (hole-hole) and electron-magnon scattering
processes into the electronic self-energy by means of the
$T$-matrix formalism, which improves the theoretical description
of the quasiparticle band structure. In particular, it is expected
to yield the correct exchange splitting in magnetic materials.

\acknowledgements Fruitful discussions with  Y.\ Mokrousov, G.\
Bihlmayer, M.\ Niesert, A.\ Gierlich, T.\ Miyake, and F.\
Aryasetiawan are gratefully acknowledged. This work was funded in
part by the EU through the Nanoquanta Network of Excellence
(NMP4-CT-2004-500198) and the European Theoretical Spectroscopy
Facility e-I3 (INFRA-2007-211956), and by the Deutsche
For\-schungs\-ge\-mein\-schaft through the Priority Programme
1145.

\end{document}